\documentclass[a4paper,aps,pra,preprintnumbers]{revtex4-2}
\usepackage[utf8]{inputenc}
\usepackage[english]{babel}
\usepackage[T1]{fontenc}
\usepackage{amssymb,amsfonts,amsmath,mathtext,enumerate,float,dsfont}
\usepackage{graphics,graphicx,epsfig,epstopdf}
\usepackage{caption}
\usepackage{cmap}
\usepackage{multirow}
\usepackage{indentfirst}
\usepackage[usenames]{color}
\usepackage{amsthm}
\usepackage{xcolor}
\usepackage{ulem}

\renewcommand{\Re}{\mathop{\mathrm{Re}}\nolimits}

\begin{document}

\title{Generation of multicomponent Schrödinger cat states in schemes with measurement of Gaussian states}

\author{E. A. Nesterova}
\author{S. B. Korolev}
\affiliation{St.Petersburg State University, Universitetskaya nab. 7/9, St.Petersburg, 199034, Russia}
\begin{abstract}
 In this work, we propose a scheme for generating superpositions of Fock states whose numbers differ by  four or eight. The proposed protocol is based on photon-number-resolving measurements on multimode Gaussian states. We compare of the generated states with multicomponent Schrödinger cat states. We evaluate the fidelity and determine how it scales with the number of detected particles. 
Furthermore, we derive the optimal configuration for the generation such states.
\end{abstract}
\vspace{10pt}
\maketitle
\section{INTRODUCTION}
Non-Gaussian quantum states have applications in various areas of quantum technology. In particular,
they play an important role in quantum error correction \cite{kor1,kor2,kor3,kor4,kor5}, quantum metrology \cite{metrology1,metrology2,metrology3},
teleportation \cite{tele2,tele3,teleport1}, and cryptography \cite{kript1,kript2}.

Among non-Gaussian states, Schrödinger cat states have attracted particular attention. Such states are used for fault-tolerant quantum computation \cite{otk1,otk2}, as well as for protecting quantum  information against particle-loss errors \cite{los1,los2,los3}.

A paradigmatic example of such states is the cat state, which is a superposition of two coherent states. In simplest form, this state can be written as:
$|\beta_2\rangle \sim  |\beta\rangle + |-\beta\rangle $, where the amplitude $ \beta \in \mathbb{C} $ determines the distance between the components in phase space \cite{Glauber1963,DODONOV1974597,PhysRevLett.57.13}. Near the origin of the phase space the Wigner function of this state exhibits an interference pattern. It has a constant extent of $\sim 1$ along one of the axes of the phase space and is limited by a scale of $\sim 1/|\beta|$ along the other \cite{Zurek2001,PhysRevA.103.053711}. This structure exhibits increased sensitivity to displacement along one of the axes of the phase space, which opens up additional possibilities for using these states in various quantum information protocols \cite{Dalvit_2006,PhysRevA.73.023803}. However, one‑dimensional scaling limits the applicability of this state in some protocols \cite{PhysRevA.103.053711}.

To address this limitation, the four-component cat state (also known as the compass state) was proposed by Zurek\cite{Zurek2001}. This state is a superposition of four coherent states with amplitude of $\beta \in \mathbb{C}$:

\begin{equation}
|\beta_4\rangle \sim|\beta\rangle + |i\beta\rangle + |-\beta\rangle + |-i\beta\rangle .
\end{equation}

Unlike the conventional cat state, the Wigner function of this state exhibits a characteristic checkerboard pattern in phase space. This structure corresponds to a two-dimensionally confined interference pattern, whose size $\sim 1/|\beta|$ can be significantly smaller than the Planck scale  \cite{PhysRevA.103.053711,DODONOV2016296} at high amplitudes. Such states provide increased sensitivity to phase-space displacements and are more robust against errors. However, their sensitivity remains limited to two perpendicular directions and is therefore anisotropic \cite{PhysRevA.108.043719,PhysRevA.106.043704}. This limitation was overcome by generalizing the four-component cat state to superpositions of six or more coherent states, which are referred to as multicomponent cat states \cite{Lan_2022}. Due to the greater distance in the Fock space, such states also offer advantages in particle-loss error correction protocols \cite{Grimsmo2020}.

However, the key problem when working with multicomponent cat is the difficulty of generating them. Such states belong to the class of non-Gaussian states \cite{deFreitas_2021}, and their deterministic generation requires high‑order nonlinear interactions \cite{Jeong_2015}. In the optical domain, such nonlinearities are difficult to realize due to the small interaction constants \cite{Jeong_2015}. At the same time, optical platforms have other advantages: low decoherence, high transmission speed, and efficient detection. For this reason, in our work we propose a probabilistic scheme for generating states close to the states of four‑component and eight‑component Schrödinger cat states.

In this work, we propose generation schemes based on photon-number-resolving measurements on multimode Gaussian states.  We investigate the states generated by the proposed schemes and compare them with multicomponent cat states in terms of their fidelity. We demonstrate that our state approximates the states of multi‑component cat states with high fidelity. We estimate the probabilities of generating such states.

\section{Generation scheme for 4-states}
\subsection{Set of generated states}

Before proceeding to the description of the states under study, we first define the Schrödinger four-component cat states. As noted earlier, the four-component cat statescan generally be defined as superpositions of coherent states of the form:

\begin{align}
    |\beta_{4}^{(\mu)}\rangle \propto |\beta\rangle+(-i)^\mu|i\beta\rangle+(-1)^\mu|-\beta\rangle +i^\mu|-i\beta\rangle, \quad \mu=0,1,2,3,
\end{align}
where the $\mu$ parameter labels one of the four possible states. In the Fock representation, these states can be expanded as an infinite superposition:

\begin{align}  \label{eq:Phi_k}
    |\beta_4^{(\mu)} \rangle \propto \sum_{j=0}^\infty \frac{\beta^{4 j+\mu}}{\sqrt{(4j+\mu)!}}|\mu+4j\rangle, \quad \mu=0,1,2,3,
\end{align}
where $|4j+\mu\rangle$ is the Fock state. In what follows, we consider states with real amplitudes $\beta\in\mathbb{R}$, and $\mu=0$. Following Zurek \cite{Zurek2001}, we refer to this state as the compass state.

\begin{figure}[H]
    \centering
    \includegraphics[width=0.5\linewidth]{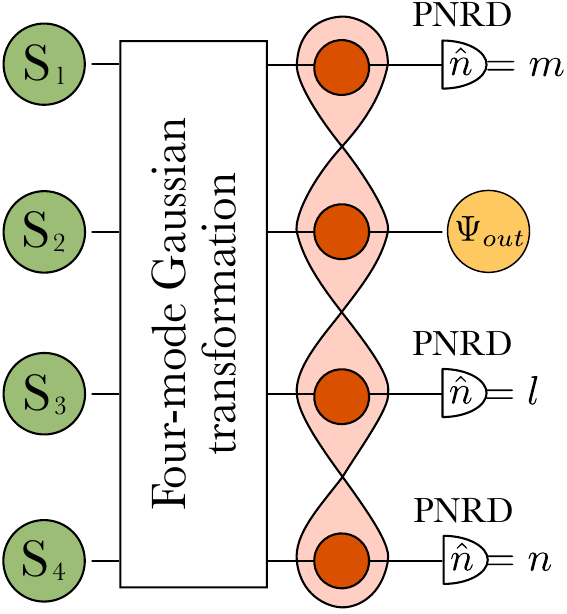}
\caption{Scheme for generating states based on the measurement of a four-mode Gaussian state. In the figure, $\text{S}_i$ denotes the $i$-th quantum oscillator prepared in a squeezed vacuum state, PNRD denotes a photon-number-resolving detector, and $\Psi_{out}$ is the generated state.}
    \label{fig:scheme_1}
\end{figure}

To generate such states, we propose a probabilistic scheme based on the measurement of multimode Gaussian states. This allows us to generate finite superpositions of Fock states whose numbers differ by multiples of four.  In what follows, for brevity, we will refer to such states as \textit{4-states}. The scheme for generating such states is presented in Fig.\ref{fig:scheme_1}.

In the proposed scheme, four independent quantum oscillators, initially prepared in squeezed vacuum states, are entangled by a four‑mode Gaussian operation. In general, the squeezing parameters may be different for different modes. At this stage, we do not specify the particular form of entangling operation. We consider the most general case. The resulting state is therefore a four‑mode Gaussian state. In the coordinate representation, its wave function is given by:
 
\begin{align} \label{gauss_state}
   \Psi_G(x_1,x_2,x_3,x_4)= \mathcal{N}_G \exp \left[-\frac{1}{2}\vec{x}^T \sigma \vec{x}\right], 
\end{align}
where $\mathcal{N}_G=\frac{(\det\Re\sigma)^{1/4}}{\pi^{N/4}}$ is the normalization factor, $\vec{x}=(x_1,x_2,x_3,x_4)^T$ is a vector of coordinates, and the matrix $\sigma$ is given as follows:

\begin{align}
    \sigma=\begin{pmatrix}
        a_1 && b_{12} && b_{13} && b_{14}\\
        b_{12} && a_2 && b_{23} && b_{24}\\
        b_{13} && b_{23} && a_3 && b_{34}\\
        b_{14} && b_{24} && b_{34} && a_4
    \end{pmatrix},
\end{align}
where $a_i, b_{ij} \in \mathbb{C}$ are parameters characterizing the Gaussian state. These parameters can take any values provided that the real part of the matrix is positive definite, $\Re(\sigma) > 0$.

In the next stage three out of four modes of the resulting Gaussian state are measured using PNRDs. Without loss of generality, we can assume that the first, third, and fourth modes are measured, with the measurement results being $m$, $l$, and $n$, respectively. As a result, at the output of the scheme, we obtain a state defined by the integral:

\begin{align}
    \Psi _{out}(x_2)\propto\int \Psi_G(x_1,x_2,x_3,x_4) \phi_n (x_1)\phi_l(x_3)\phi_m (x_4) d x_1 dx_3 dx_4,
\end{align}
where $\phi _{j}(x)=\frac{ e^{-\frac{x^2}{2}} H_j(x)}{\sqrt[4]{\pi } \sqrt{2^{j}\, j!}}$ is the wave function of the $j$-th Fock state in the coordinate representation, and $H_{j}(x)=(-1)^{j}e^{x^{2}}{\frac {d^{j}}{dx^{j}}}e^{-x^{2}}$ is the Hermite polynomial.

To bring the generated state closer to the compass state, we need to adjust the parameters of the Gaussian state so that the output state is a 4‑state. We have found the most general solution that allows us to generate stateswhose wave functions are given by

\begin{multline} \label{out_state}
    |\Psi^{out}_{n,m}\rangle \propto   \sum _{j=\max \left(0,\left\lceil \frac{1}{4} (2 n-m)\right\rceil \right)}^n 
   \left(-\frac{c_2^4}{4c_3^2}\right)^j \frac{(2 j)!(2n- 2j)!}{\sqrt{\left(m-2 n+4 j\right)!}} \binom{n}{n-j}\\
   \,
   _2\tilde{F}_1\left(-m,-m+2 n-4 j;-m+2 n-2 j+1;\frac{c_1 c_3}{c_2 c_4}\right)
   |m-2 n+4 j\rangle.
\end{multline}
Here, $c_i \in \mathbb{C}$ are new variables that characterize the generated state. The general solution for generating such a class of states can be written as follows: $l=n$. The parameters of the Gaussian state are functions of the generated-state parameters: $a_i=f_i(c_1,c_2,c_3,c_4,c_5)$, $b_{ij}=f_{ij}(c_1,c_2,c_3,c_4,c_5)$ (the explicit dependence of the Gaussian state parameters on the new variables of the generated state $c_i$ is given in Appendix \ref{append_solut}). The found solution links the ten input parameters of the Gaussian state [Eq. (\ref{gauss_state})] with the five parameters $c_i$ of the generated state. Knowing the parameters of the generated state $c_i$, we unambiguously determine the parameters of the Gaussian state required to generate such a state. Consequently, when the same number of particles is measured at the third and fourth detectors, and the parameters of the four‑mode Gaussian state areappropriately chosen, we generate a 4‑state in the scheme.

From the explicit form of the state vector (\ref{out_state}), it follows that the generated state is a finite superposition of Fock states whose numbers differ by four. This is the same structure as that of a four‑component Schrödinger cat in the Fock representation (\ref{eq:Phi_k}). The choice of one of the four possible states of the four‑component Schrödinger cat in our scheme is determined by the smallest number of the Fock state appearing in the superposition (\ref{out_state}). This number depends on the measurement outcomes at the detectors (on the numbers $m$ and $n$). For example, the measurement yields numbers lying in the set: $\lbrace n,m\rbrace=\lbrace p+q,2(p-q)\rbrace _{q=0}^p$, where $p$ is a positive integer, the output state has the form: $c_0|0\rangle+c_4|4\rangle+\dots c_{4p}|4p\rangle$. That is, the state is similar to the compass state.

To ensure that the generated states are close to the compass state, the superposition coefficients must be chosen appropriately. In the present scheme, the coefficients depend on two parameter ratios: $-c_2^4/c_3^2$ and $c_1 c_3/(c_2 c_4)$. Thus, out of the five parameters that remain free after fixing the parameters of the Gaussian state, only two parameter ratios determine the state itself. The remaining parameters can therefore be used to simplify the generation scheme and maximize the probability of obtaining the required states. We will discuss this in more detail below.

\subsection{Generation scheme}
\begin{figure}[H]
    \centering
    \includegraphics[width=0.7\linewidth]{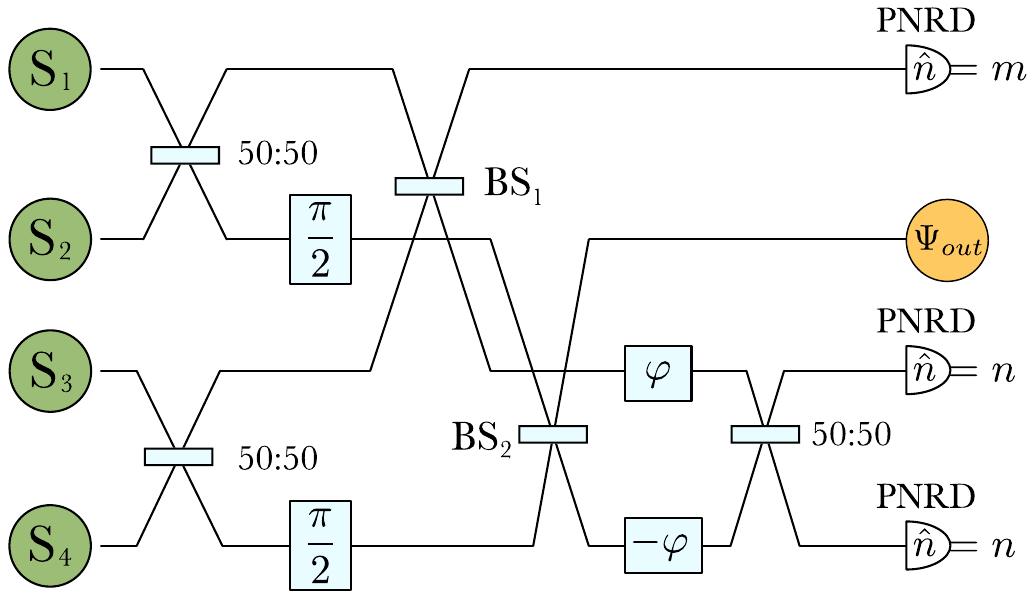}
  \caption{Scheme for generating 4-states. In the figure, $\text{S}_1$ and $\text{S}_2$ are two orthogonally squeezed vacuum states with squeezing parameter $r_1 \in \mathbb{R}$, $S_3$ and $S_4$ are two orthogonally squeezed vacuum states with squeezing parameter $r_2 \in \mathbb{R}$, $\text{BS}_i$ are beam splitters with transmittance $t_i$, the blue rectangles denote phase shifters, PNRD stands for a photon-number-resolving detector, and $\Psi_{out}$ is the generated state.}
    \label{fig:scheme_2}
\end{figure}

In the previous subsection, we presented the most general scheme for generating a four-mode Gaussian state. Let us now specialize this scheme to the generation of states similar to compass states. To do this, let us introduce a constraint on the parameters of the generated state. Comparing the state vectors (\ref{eq:Phi_k}) and (\ref{out_state}), we see that for a compass state, the superposition coefficients are positive real numbers. In order for the coefficients of the generated state to satisfy the same conditions, we need, at least require the reality and positivity of $-c_2^4/c_3^2 \equiv A>0$, as well as the reality of the ratio $c_1 c_3/(c_2 c_4) \equiv B\in\mathbb{R}$. Furthermore, to simplify the generation scheme, we will also choose $c_5=-1$ and $c_4=\frac{\left(c_1^2-4\right) c_3}{c_1 c_2}$. Taking these constraints into account, together with the relations (\ref{a1})-(\ref{b34}), we can construct the covariance matrix of the required Gaussian state, perform its symplectic decomposition, and identify the most general scheme for generating such a Gaussian state. The resulting scheme for generating 4-states is shown in Fig.\ref{fig:scheme_2}

\noindent In the scheme, two pairs of orthogonally squeezed states $\hat{\mathcal{S}}(r_1)|0\rangle_1$, $\hat{\mathcal{S}}(-r_1)|0\rangle_2$ and $\hat{\mathcal{S}}(r_2)|0\rangle_3$, $\hat{\mathcal{S}}(-r_2)|0\rangle_4$, with real squeezing parameters, are mixed pairwise on balanced beam splitters. As a result, two pairs of two-mode squeezed vacuum states are obtained. They are subsequently mixed with each other and measured using PNRDs. The remaining unmeasured mode is then prepared in a 4-state. Moreover, the superposition coefficients are real. 

The presented scheme contains five real parameters $\lbrace r_1,r_2, t_1,t_2,\varphi \rbrace$, which can be related to the parameters $c_1,\dots,c_5$. This scheme is not general, as it was obtained under the constraint that $c_5=-1$ and $c_4=\frac{\left(c_1^2-4\right) c_3}{c_1 c_2}$. In theory, it is possible to construct a general scheme that depends on an even larger number of parameters. However, in any such scheme, the set of generated states coincides with that obtained in the scheme shown in Fig. \ref{fig:scheme_2}. This is because the generated states depend on only two parameters, $A$ and $B$. Therefore, without loss of generality, we restrict our analysis to the states generated in the scheme presented in Fig. \ref{fig:scheme_2}.

\section{Comparison with compass states}
\subsection{Fidelity bound for compass states}
Before comparing the states generated in the considered schemes with the compass state, we first determine how closely the compass state can be approximated in theory.  As a numerical measure for comparing the two states $|\beta _4^{(0)}\rangle$ and $|\Psi_4\rangle$, we use the fidelity, which has the following form:

\begin{align}
    \mathcal{F}=\left|\langle \beta _4^{(0)}|\Psi_4 \rangle \right|^2.
\end{align}

Our goal is to determine the upper bound on the fidelity between the compass state and the 4-state $|\Psi_4 \rangle =\sum _{j=0}^p \alpha_j |4j\rangle$. We refer to a state achieving this upper bound as the optimal state. The main difference between this state and the state under study (\ref{out_state}) is that here the superposition coefficients are limited only by the normalization condition $\sum_{j=0}^p |\alpha_j|^2=1$. As the number $p$ increases, so does the number of parameters of the optimal state. For the state (\ref{out_state}), the superposition coefficients always depend on only two parameters, which restricts the class of states that can be generated.

Using the Cauchy–Bunyakovsky–Schwarz inequality, one can determine the upper bound of the fidelity for generating an arbitrary state and to select the superposition coefficients of the optimal state. Using these results, we estimated the maximum achievable fidelity between the optimal and the compass states:
\begin{align} \label{fid_limit}
    \mathcal{F}^{max}_4 (\beta,p)=\sum _{j=0}^p \frac{2 \beta ^{8 j}}{{(4 j)! \left(\cos \left(\beta ^2\right)+\cosh
   \left(\beta ^2\right)\right)}}.
\end{align}
The obtained value depends on two parameters: the amplitude $\beta$ of the compass state, as well as the number of superposition terms $p$ in the optimal state. Note that in the limit $p\rightarrow \infty$, the maximum fidelity tends to one, which indicates the correctness of the estimate. This expression allows us to assess how close the fidelity of the compass states generated in our scheme approach the maximum achievable fidelity.

\subsection{Fidelity of compass states in our scheme}
\subsubsection{Superpositions with the vacuum state}
Let us now examine the fidelity of the compass states obtained in the scheme shown in Fig. \ref{fig:scheme_2}. For each amplitude $\beta$, we maximize the fidelity by optimizing the parameters $A$ and $B$.  In other words, we optimize the parameters of the generated state to approximate the corresponding compass state. As a result, we obtain a fidelity:

\begin{align} \label{eq:fidelity}
    \mathcal{F}_{n,m}(\beta)= \max_{A,B} \left[\left|\langle \Psi^{out}_{n,m}|\beta_4 ^{(0)} \rangle \right|^2\right],
\end{align}
 that depends on the amplitude $\beta$ and on the numbers of detected particles $n$ and $m$.

As noted above, the numbers of detected particles affect both the minimum and maximum Fock-state numbers appearing in the superposition (\ref{out_state}). Moreover, different detection results can lead to states with identical Fock-state decompositions. It is therefore inconvenient to characterize the generated states solely by the detected particle numbers. Instead, we characterize the fidelity by two parameters: the minimum Fock-state index present in the superposition and the total number of terms in the decomposition. 

Let us start the analysis with the case where the superposition state (\ref{out_state}) contains the vacuum state $|0\rangle$. As noted earlier, the generated state takes the form $c_0|0\rangle+c_4|4\rangle+\dots c_{4p}|4p\rangle$, when the measured particle numbers belong to the set: $\lbrace n,m\rbrace=\lbrace p+q,2(p-q)\rbrace _{q=0}^p$. For a fixed number of superposition terms $p$, there are several pairs of detection outcomes that generate states containing the same set of Fock states. The only difference lies in the superposition coefficients, which, as follows from expression (\ref{out_state}), depend on the specific numbers $n$ and $m$. Taking this into account, maximize the fidelity not only over the parameters $A$ and $B,$ but also over the detection outcomes $n$ and $m$. As a result, the fidelity of generating compass states in this case is given by:

\begin{align} \label{fid_k0}
    \mathcal{F}(\beta,p)= \max_{q={0,\dots,p}} \left[\max_{A,B} \left[\left|\langle |\Psi^{out}_{p+q,2(p-q)}|\beta_4^{(0)} \rangle \right|^2\right]\right],
\end{align}

which depends on the amplitude $\beta$ and on the number of terms in the superposition $p$.

\begin{figure}[H]
    \centering
    \includegraphics[width=0.5\linewidth]{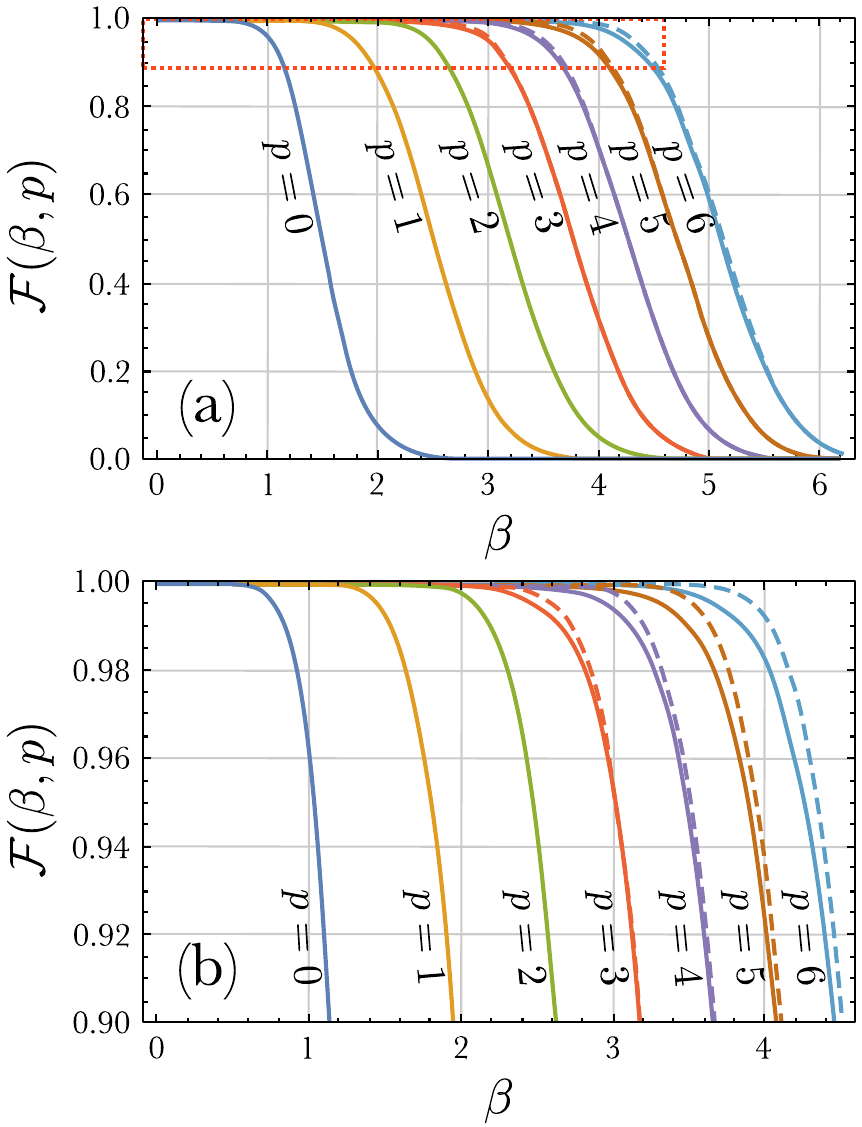}
    \caption{(a) Fidelity $\mathcal{F}(\beta,p)$ (solid curves) and $\mathcal{F}^{max}_4(\beta,p)$ (dashed curves) as a function of amplitude $\beta$ for different numbers of superposition terms $p$. (b) Enlarged image of the region  indicated by the red dashed rectangle.}
    \label{fig:fid_comp_0}
\end{figure}

Fig. \ref{fig:fid_comp_0} shows the fidelity (\ref{fid_k0}) as a function of the amplitude  $\beta$ for different numbers of superposition terms $p$. For comparison, the figure also shows the corresponding dependence for the maximum achievable fidelity (\ref{fid_limit}).

\noindent The figure shows that as the number of terms in the superposition increases, the fidelity approaches unity. In other words, increasing the number of terms broadens the range of amplitudes over which high-fidelity compass states can be generated. The optimal choice of detection outcomes is $n=p$ and $m=2p$. This choice maximizes the fidelity of the generated compass states. Thus, by detecting $n=5$ and $m=10$  particles, one can generate a compass state with amplitude $\beta=3.5$and fidelity $\mathcal{F}(3.5,5)=0.99$ Detecting  $n=6$ and $m=12$ particles yields a state with $\beta=3.9$ and the same fidelity.

Let us compare the fidelity of the compass states generated in the scheme presented in Fig. \ref{fig:scheme_2} with the maximum achievable fidelity. For a small number of superposition terms, the fidelity is the same in both cases. However, as $p$ increases, the difference becomes increasingly significant. This is because our scheme has only two controllable parameters (\ref{out_state}) that affect the generated state, whereas the optimal state has
$p$ such parameters. When $p\leq2$, the two schemes have the same number of controlled parameters, so the fidelity coincides. For $p=3$, the number of parameters in the optimal state exceeds the number of parameters in our state, and the fidelity in the real case falls below the limiting value. Nevertheless, even at $p=6$, the maximum difference between the two cases considered is approximately one percent.

Thus, our scheme allows for the generation of compass states with fidelities comparable to the theoretical limit.

\subsubsection{Superpositions without the vacuum state}
A distinctive feature of our scheme is that, for certain detection outcomes, we can generate states in which the minimum Fock state present in the decomposition is not the vacuum state. Let us investigate how closely such states can approximate compass states.

\begin{figure}[H]
    \centering
    \includegraphics[width=0.5\linewidth]{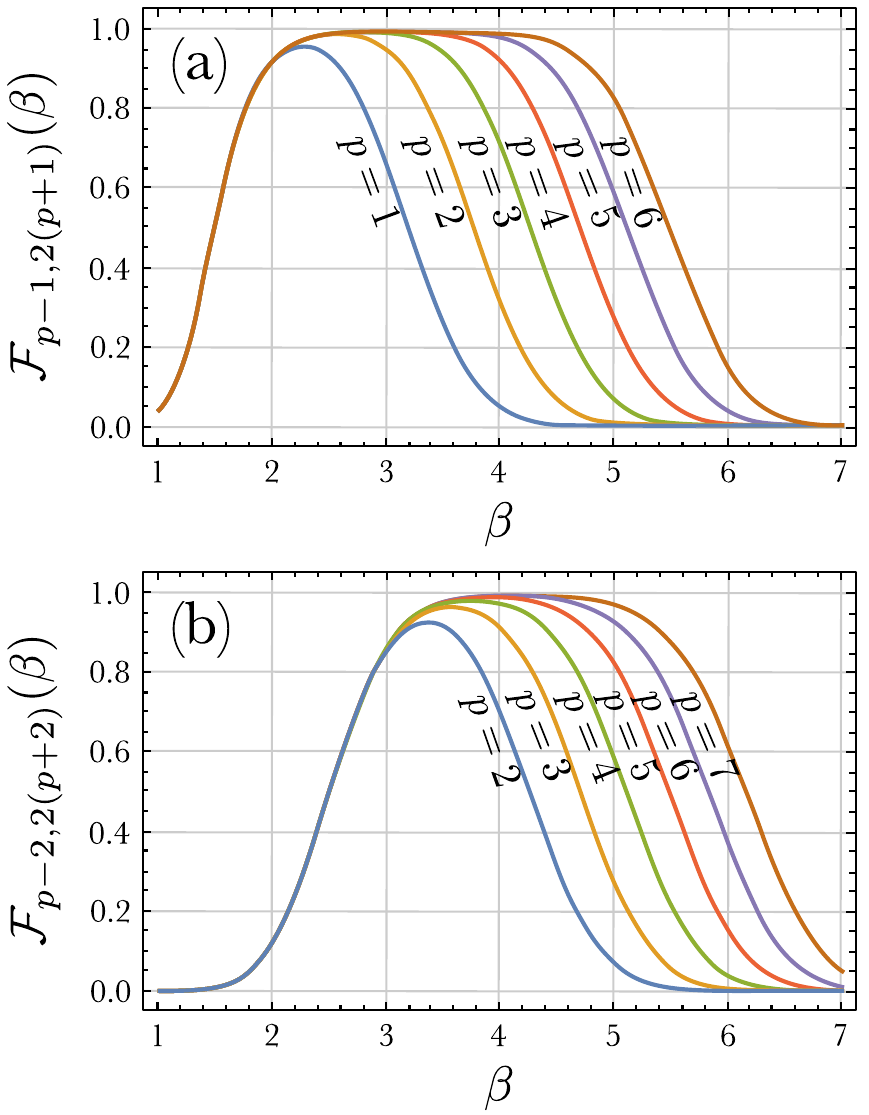}
    \caption{(a) Fidelity between the compass state and the state starting from the fourth Fock state as a function of $\beta$ for different numbers of superposition terms. (b) Fidelity between the compass state and the state starting from the eighth Fock state as a function of $\beta$ for different numbers of superposition terms.}
    \label{fig:start4-8}
\end{figure}

Unlike states containing a vacuum state, a state without a vacuum corresponds to a unique pair of detection outcomes. For example, detecting $n=p-1$ and $m=2(p+1)$ particles, with $p\geqslant 1$, yields the state $c_4 |4\rangle+c_8 |8\rangle+\cdots+c_p|4p\rangle$, which starts with the fourth Fock state. Detecting $n=p-2$ and $m=2(p+2)$, particles, with$p\geqslant 2$, yields states of the form $c_8 |8\rangle+c_12 |12\rangle+\cdots+c_p|4p\rangle$, which start from the eighth Fock state. Let us determine the maximum fidelity of these states with the compass state, $\mathcal{F}_{n,m}(\beta)$. Fig. \ref{fig:start4-8} shows the fidelity as a function of $\beta$ for different numbers of measured particles (different numbers of superposition terms $p$).

\noindent The graphs show that in both cases, increasing the number of superposition terms $p$ broadens the range of amplitudes for which high-fidelity compass states can be generated. Comparing the graphs shows that the maximum of the fidelity curve shifts to larger amplitudes as the minimum Fock-state number of the superposition increases. This is because, as the amplitude of the compass state increases, the contribution of lower Fock states in the expansion (\ref{eq:Phi_k}) decreases, while the contribution of higher states grows. Appendix \ref{append_distrib_coef} presents the distributions of the expansion coefficients of the compass state (\ref{eq:Phi_k}) over Fock states for various amplitudes.

\begin{figure}[H]
    \centering
     \includegraphics[width=0.5\linewidth]{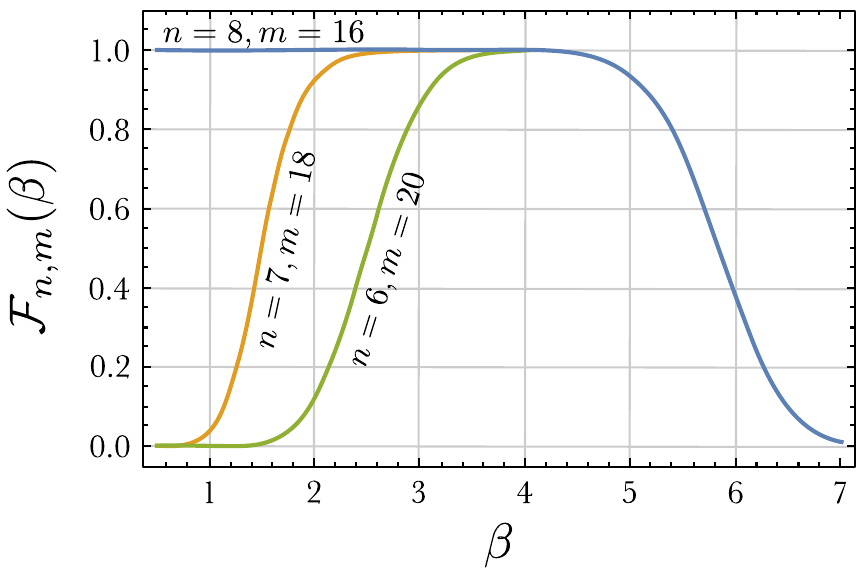}
     \caption{Fidelity of the compass state as a function of amplitude for three cases: the superposition starts from the vacuum state (blue, $n=8$, $m=16$), from the fourth Fock state (orange, $n=7$, $m=18$), and from the eighth Fock state (green, $n=6$, $m=20$).}
    \label{fig:n1_dependence}
\end{figure}

To illustrate this, we plot the fidelity as a function of the amplitude for three cases: the minimum Fock-state number is
$0$ (generated by detecting $n=8$ and $m=16$), $4$ (generated by detecting $n=7$ and $m=18$), and $8$ (generated by detecting $n=6$ and $m=20$). In all three cases, the maximum Fock-state number is fixed at $32$. The corresponding fidelity curves are shown in Fig. \ref{fig:n1_dependence}.

\noindent The case starting from vacuum state has the widest range of amplitudes with high fidelity. Changing the minimum Fock-state number shifts the high-fidelity range to larger amplitudes, but the right boundary is the same for all curves. This suggests that the right boundary is determined by the maximum Fock-state number ($|4p\rangle$) present in the decomposition.

Generating superposition states that start from a higher Fock state requires detecting a larger number of particles. For a fixed maximum Fock-state index $4p$, the required number of detected particles depends on the starting state: $m = 2p$ for the vacuum state, $m = 2(p+1)$ for the state starting from the fourth Fock state, and $m = 2(p+2)$ for the state starting from the eighth Fock state. Since it is challenging to construct detectors that can resolve a large number of photons in practice~\cite{app142311249}, the strategy based on generating superpositions starting from the vacuum state is the most promising from the experimental point of view.

\subsection{Generation probability of compass states}
The generation scheme under consideration is probabilistic, as successful generation requires the detection of a certain number of particles, with the generation probability given by the probability of the corresponding detection event. For the scheme shown in Fig. \ref{fig:scheme_2}, the generation probability is given by: 
\begin{multline} \label{prob_gen-4}
    \mathcal{P}_{n,m}(A,B,c_2)=
    \frac{\pi^2 \mathcal{N}_G^2}{m! (n!)^2}\left|\frac{\left((B-1) c_2^2+2\right) \left(A B-2 c_2^2\right)}{2 A (B-1) B} \left(\frac{1}{(B-1) B}\right)^m \left(\frac{c_2^2}{A}\right)^{2n} \right|\\
   \times \sum _{j=\max \left(0,\left\lceil \frac{1}{4} (2 n-m)\right\rceil \right)}^n\left(\left(\frac{A}{4}\right)^{j}\frac{ (2 j)!\,(2n-2j)!}{\sqrt{(m-2n+4j)!}} \binom{n}{n-j} \, _2\tilde{F}_1(-m,-m+2 n-4 j;-m+2 n-2 j+1;B)\right)^2.
\end{multline}

The probability depends on the parameters  $A$ and $B$ of the generated state, and the parameter $c_2$, which does not affect the generated state. The normalization factor of the Gaussian state $\mathcal{N}_G$ also depends on these three parameters. This means that the parameter $c_2$ is a free parameter that can be optimized to maximize the generation probability \cite{Korolev_fgen,Korolev_2024,Fiurasek_2026_1,Fiurasek_2026_2}. Knowing these parameters, one can determine the parameters of the generation scheme that maximize the probability of generating states close to the compass states for given detection outcomes $n$ and $m$.

To calculate the probability of generating compass states, we use Eq. (\ref{prob_gen-4}) with the values of 
$A$ and $B$ obtained from the fidelity optimization. We then maximize the probability over the free parameter $c_2$.

\begin{figure}[H]
    \centering
    \includegraphics[width=0.5\linewidth]{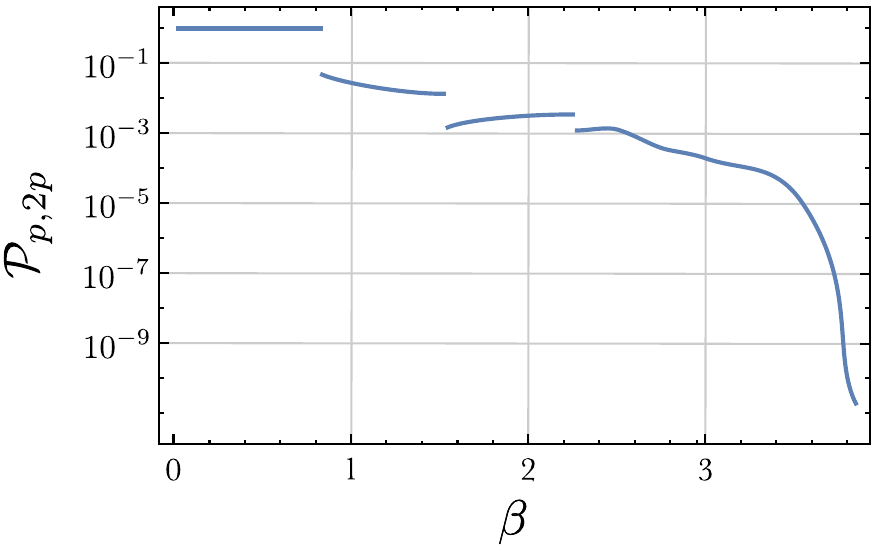}
    \caption{Maximum probability of generating compass states with fidelity at least $0.99$ as a function of amplitude.}
    \label{fig:prob}
\end{figure}

We consider the probability of generating states in the scheme for the detection outcomes $n=p$ and $m=2p$ particles. Fig. \ref{fig:prob} shows the maximum probability of generating compass states with fidelity  at least $0.99$ as a function of their amplitude.

\noindent 
The discontinuities in the graph arise because, in different amplitude ranges, states with maximum fidelity are generated at different values of $p$. As the amplitude of the generated state increases, the probability of obtaining it decreases. This is because generating states with large amplitudes requires detecting a large number of particles on three detectors, which is an unlikely event. The figure shows that the maximum probability of $100\, \% $ is achieved for compass states with amplitude $0<\beta<0.89$. In this amplitude range, the compass state is close to the vacuum state, which can be generated deterministically in our scheme. As the amplitude increases, the generation probability decreases. For example, the probability of generating a state with amplitude  $\beta \approx 1.5$ is about $1.4 \% $. A further increase in amplitude leads to a sharp drop in generation probability to values on the order of $10^{-10}$. It is worth noting that even such probabilities are experimentally observable. At source repetition rates on the order of several GHz, even probabilities as low as \(10^{-10}\) would correspond to characteristic generation times of only a few seconds.

\section{Generation of eight-component Schrödinger cat states}
\subsection{Set of generated states}
In the previous section, we showed that the proposed scheme allows us to generate four-component Schrödinger cat states with high fidelity. Let us now consider the generation of eight-component Schrödinger cat states. By definition, such a state is given by a superposition of coherent states and has the following decomposition in the Fock-state basis:

\begin{equation}\label{8cat}
|\beta_8 ^{(\mu)}\rangle  \propto \sum_{j=0}^{7} e^{-i \pi \mu j/4} |e^{i\pi j/4}\beta\rangle
 \propto \sum_{j=0}^{\infty} \frac{\beta^{8j+\mu}}{\sqrt{(8j+\mu)!}} \, |8j+\mu\rangle, \quad \mu=0,1,2,\dots, 7.
\end{equation}
where $\beta$ is the amplitude of the coherent state. Such a state contains only Fock states whose numbers differ by multiples of eight.

\begin{figure}[H]
    \centering
    \includegraphics[width=0.4\linewidth]{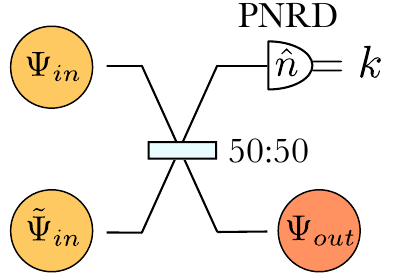}
    \caption{Scheme for generating 8-states. In the figure, $\Psi_{in}$ and $\tilde{\Psi}_{in}$ are two input 4-states, the 50:50 block denotes a balanced beam splitter, PNRD stands for a photon-number-resolving detector, and $|\Psi^{out}\rangle$ is the output state.}
    \label{fig:scheme_3}
\end{figure}

To generate states close to an eight-component cat, we need to modify the scheme proposed earlier so that the output is a finite superposition of Fock states whose numbers differ by multiples of eight. We will subsequently refer to such states as \textit{8‑states}. To generate these states, we use the scheme presented in Fig.\ref{fig:scheme_3}

\noindent In the scheme, two 4-states are mixed on a balanced beam splitter, after which one mode is measured by a PNRD. For definiteness, we assume that the detector registered $k$ particles. The remaining unmeasured mode will be in the output state.

We use as input states the states obtained in the scheme of Fig. \ref{fig:scheme_2}.  One input state is $|\Psi^{in}_1 \rangle=|\Psi^{out}_{n,m}\rangle$, defined by expression (\ref{out_state}). It depends on the parameters $A$ and $B$ and the detection outcomes $n$ and $m$. The other input state is the same state, but rotated in phase space by $\pi/4$: $|\Psi^{in}_2 \rangle=e^{i\pi \hat{n}/4}|\Psi^{out}_{n,m}\rangle$. In optics, such a rotation can be achieved using a phase plate of a certain thickness.

After mixing on a balanced beam splitter and detecting $k$ photons in one of the modes, we obtain an 8-state of the form:

\begin{align} \label{out_state_8}
  |\Psi_{n,m,k}^{out}\rangle \propto  2^{2 n-m}\sqrt{k!}\begin{cases}
           \sum \limits_{j=0}^{\left\lfloor \frac{1}{8} (2 (m+2 n)-k)\right\rfloor}  \gamma_{2j}(A,B ;n,m,k) |2 m+4 n-k-8 j\rangle, \quad \text{if}\quad k\, \text{-- even}\\
        \sum \limits_{j=0}^{\left\lfloor \frac{1}{8} (2 (m+2 n-2)-k)\right\rfloor}  \gamma_{2j+1}(A,B;n,m,k)|2 m+4 n-k-4 (2 j+1)\rangle, \quad \text{otherwise}
    \end{cases},
\end{align}
where the expansion coefficients $\gamma_{2r+1}(A,B)$ are defined in Appendix \ref{append_decompos_coef_8}. The obtained expansion shows that the expansion coefficients of the generated states depend on the input state parameters $A$, $B$, $n$, and $m$, as well as on the number of measured particles $k$. By selecting these parameters and the detection outcomes, we obtain states from the set (\ref{out_state_8}).

Furthermore, as before, the number of terms in the superposition and the minimum Fock state number are by the parameters $n$ and $m$ of the input states and by the number of detected particles $k$. By selecting these parameters, we can obtain an approximation to one of the eight states of the eight‑component Schrödinger cat (\ref{8cat}).

\subsection{Fidelity of eight-component Schrödinger cat states}

Let us assess how close the generated states are to the states of an eight‑component Schrödinger cat. As before, we consider generating a state with $\mu=0$ and $\beta \in \mathbb{R}$. We found that states whose decomposition includes the vacuum state provide the best approximation to the four‑component cat state. Therefore, we will use the same type of states to approximate the eight‑component cat state.

Let us compare the obtained states with the states of an eight‑component cat.
We determine the maximum fidelity between the generated state and an eight-component cat state: $\tilde{\mathcal{F}}(\beta,p)= \max_{n,m,k} \left[\max_{A,B} \left[\left|\langle |\Psi^{out}_{n,m,k}|\beta_8^{(0)} \rangle \right|^2\right]\right]$. We optimize the fidelity with respect to the state parameters $A$ and $B$ and the detection outcomes $n$, $m$, and $k$. As a result, the fidelity depends on the state amplitude $\beta$ and on the number of superposition terms $p$.

Just as in the case of four‑component cat states, let us determine the maximum fidelity between the state of an eight‑component Schrödinger cat with $\mu=0$ and the optimal state $|\Psi_8\rangle=\sum _{j=0}^p \alpha_{j}|8j\rangle$, which is an 8‑state. This value is given by the following expression:
\begin{align}
    \tilde{\mathcal{F}}^{max}_8 (\beta,p)=|\langle \beta _8^{(0)}|\Psi _8\rangle|^2=\sum_{j=0}^{p} \frac{4 \beta ^{16 j}}{(8 j)! \left(\cos \left(\beta ^2\right)+\cosh \left(\beta ^2\right)+2 \cos \left(\frac{\beta
   ^2}{\sqrt{2}}\right) \cosh \left(\frac{\beta ^2}{\sqrt{2}}\right)\right)}.
\end{align}

\begin{figure}[H]
    \centering
    \includegraphics[width=0.5\linewidth]{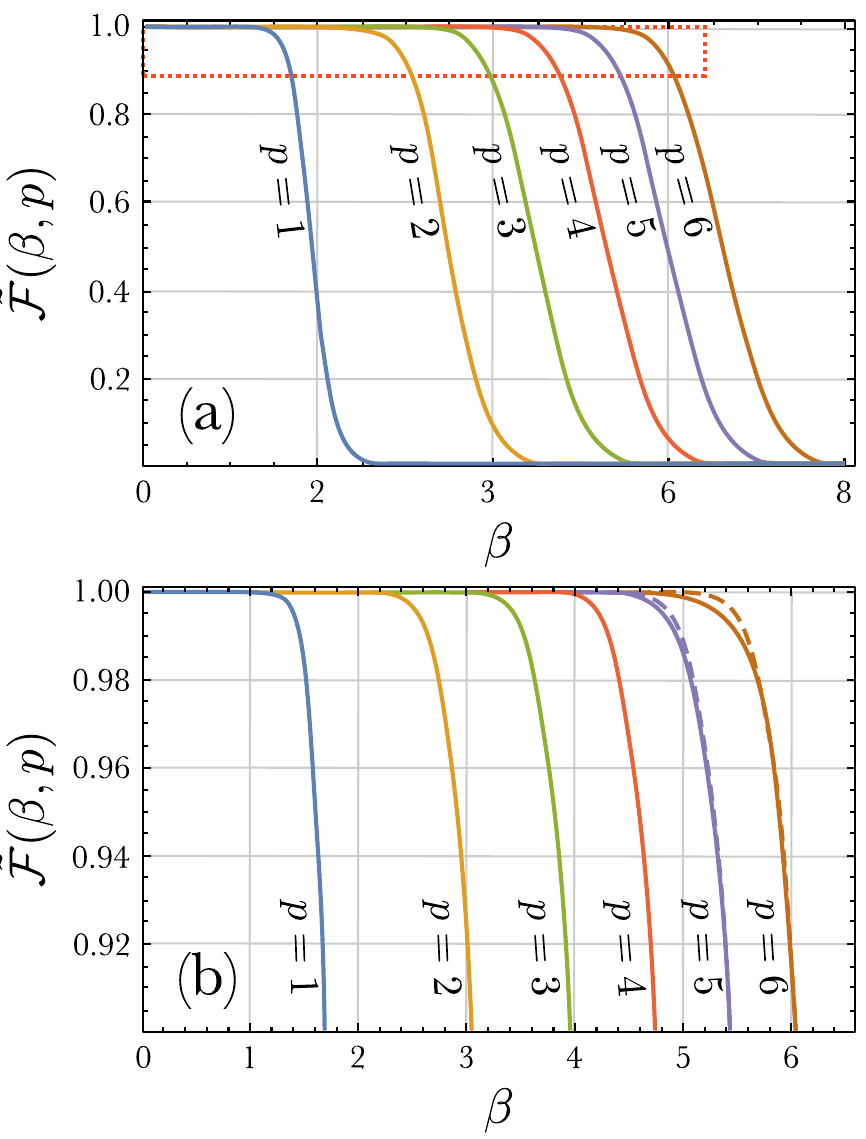}
    \caption{(a) Fidelity $\tilde{\mathcal{F}}(\beta,p)$ (solid curves) and $\tilde{\mathcal{F}}^{max}_8(\beta,p)$ (dashed curves) as a function of amplitude $\beta$ for different numbers of superposition terms $p$. (b) Zoomed-in region indicated by the red dashed rectangle.}
    \label{fig:8cat}
\end{figure}

Let us now determine the fidelity of the generated state relative to the theoretically achievable maximum. Fig. \ref{fig:8cat} shows the fidelity $\tilde{\mathcal{F}}(\beta,p)$ and $\tilde{\mathcal{F}}^{max}_8 (\beta,p)$as a function of $\beta$ 
for different numbers of superposition terms $p$

\noindent The overall trend of the graph mirrors the behavior observed when studying compass states. As the number of terms in the superposition increases, the fidelity of the generated states increases, and the range of amplitudes over which $\mathcal{F} > 0.99$ expands. As in the case of the compass state, for a small number of superposition terms the fidelities coincide, but as the number of terms increases ($p\geqslant 5$), a small difference appears (less than 1\%).  This confirms that the considered scheme makes it possible to generate eight-component cat states with fidelity close to the theoretical limit

Maximum fidelity is achieved for $k=0$ and the same detection outcomes as for the compass states: $n = p$, $m = 2p$. Thus, when $p=1$ ($n=1$, $m=2$, $k=0$), the maximum amplitude achieved with a fidelity of $0.99$ is $\beta \approx 1.46$. When $p=2$, it increases to $\beta \approx 2.63$, and when $p=6$, it increases to $\beta \approx 5.59$.

The only drawback is the low probability of generating eight‑component cat states. This probability can be bounded from above by the square of the probability of generating a four‑component cat state:$P_{8cat} \leqslant (P_{n,m}(A,B,c_2))^2$. Thus, the probability of generating an eight‑component cat with amplitude $\beta \approx 4$ is below $10^{-20}$.  If the amplitude is reduced to $\beta \approx 3$, the probability is approximately $10^{-10}$, which is experimentally achievable at current squeezed-state generation rates.

Thus, using two 4‑states and the scheme shown in Fig. 2, it is possible to generate 8‑states. Moreover, this approach can be generalized. By using 8-states as the inputs to the presented scheme, it is possible to generate 16‑states, and so on. The generated states can be made closer to multi‑component cat states by tuning the state parameters.However, with each such step, the probability of generating a state decreases quadratically. Therefore, in practical quantum information protocols, one must find a compromise between the properties of the states and the experimental probability of obtaining them.

\section{CONCLUSION}
In this paper, we propose a probabilistic scheme for generating non‑Gaussian states that are close to multicomponent Schrödinger cat states, based on measuring the number of particles in the modes of multimode Gaussian states.

It was shown that measuring three modes of a four‑mode Gaussian state allows one to generate finite superpositions of Fock states whose photon numbers differ by multiples of four (4‑states). We identified the conditions on the parameters of the initial Gaussian state that ensure the output state belongs to this class. We also constructed a specific optical implementation of the scheme based on squeezed vacuum states, beam splitters, phase shifters, and PNRD detectors.

We compared the generated 4‑states with one of the four‑component Schrödinger cat states (the compass state). We showed that increasing the number of superposition terms broadens the range of amplitudes for which the generated states approximate the compass state with high fidelity. In addition, we established the maximum achievable fidelity for generating compass states in schemes based on particle‑number measurements on multimode Gaussian states. We have shown that the proposed scheme can generate compass states with fidelity close to the theoretical limit.

We also examined 4-states in which the superposition does not start with the vacuum state but with a higher Fock state. We showed that excluding lower Fock states shifts the high-fidelity region to larger amplitudes. However, this strategy requires detecting a larger number of particles and, consequently, has a lower generation probability and is more challenging to implement. This makes superpositions containing the vacuum state the most promising approach for for generating four‑component Schrödinger cat states.

In addition, we estimated the probability of generating states close to the compass state in our scheme. We showed that the probability decreases rapidly as the amplitude of the generated state increases.

Another important result of the work is the proposed generation scheme, in which, using two 4‑states, it is possible to obtain a final superposition of Fock states whose numbers differ by a multiple of eight (8‑states). We demonstrated that, with an optimal choice of parameters, such states can also approximate eight‑component Schrödinger cat states with high fidelity. At the same time, the difference from the maximum achievable fidelity remains less than one percent for a sufficiently large number of superposition components.

The proposed generation scheme allows for a natural recursive generalization: by using 8‑states as input states, one can construct a generation scheme for 16‑states, and so on. However, the probability of successful generation decreases quadratically at each subsequent step. Therefore, in real quantum information protocols, it is important to take into account the trade‑off between the characteristics of the states and the probabilities of generating them.

This research was supported by the Theoretical Physics and Mathematics Advancement Foundation "BASIS" (Grant No. 24-1-3-14-1).

\bibliography{bibliography}

@misc{metrology1,
      title={Conclusive nonlinear phase sensitivity limit for a Mach-Zehnder interferometer with single-mode non-vacuum inputs}, 
      author={Jian-Dong Zhang and Zi-Jing Zhang and Jun-Yan Hu and Long-Zhu Cen and Yi-Fei Sun and Chen-Fei Jin and Yuan Zhao},
      year={2019},
      eprint={1906.10867},
      archivePrefix={arXiv},
      primaryClass={quant-ph},
      url={https://arxiv.org/abs/1906.10867}, 
}

@article{Grimsmo2020,
  title = {Quantum Computing with Rotation-Symmetric Bosonic Codes},
  author = {Grimsmo, Arne L. and Combes, Joshua and Baragiola, Ben Q.},
  journal = {Phys. Rev. X},
  volume = {10},
  issue = {1},
  pages = {011058},
  numpages = {32},
  year = {2020},
  month = {Mar},
  publisher = {American Physical Society},
  doi = {10.1103/PhysRevX.10.011058},
  url = {https://link.aps.org/doi/10.1103/PhysRevX.10.011058}
}

@article{metrology2,
doi = {10.1088/1674-1056/28/4/044203},
url = {https://doi.org/10.1088/1674-1056/28/4/044203},
year = {2019},
month = {apr},
publisher = {Chinese Physical Society and IOP Publishing Ltd},
volume = {28},
number = {4},
pages = {044203},
author = {Hou, Li-Li and Sui, Yong-Xing and Wang, Shuai and Xu, Xue-Fen},
title = {Quantum interferometry via a coherent state mixed with a squeezed number state*},
journal = {Chinese Physics B}
}

@article{metrology3,
  title = {Quantum metrology of two-photon absorption},
  author = {S\'anchez Mu\~noz, Carlos and Frascella, Gaetano and Schlawin, Frank},
  journal = {Phys. Rev. Res.},
  volume = {3},
  issue = {3},
  pages = {033250},
  numpages = {9},
  year = {2021},
  month = {Sep},
  publisher = {American Physical Society},
  doi = {10.1103/PhysRevResearch.3.033250},
  url = {https://link.aps.org/doi/10.1103/PhysRevResearch.3.033250}
}

@article{teleport1,
    author = "Opatrny, T. and Kurizki, G. and Welsch, D. -G.",
    title = "{Continuous-variable teleportation improvement by photon subtraction via conditional measurement}",
    eprint = "quant-ph/9907048",
    archivePrefix = "arXiv",
    doi = "10.1103/PhysRevA.61.032302",
    journal = "Phys. Rev. A",
    volume = "61",
    pages = "032302",
    year = "2000"
}

@article{tele2,
    author = "Zinatullin, E. R. and Korolev, S. B. and Golubeva, T. Yu.",
    title = "{Teleportation protocols with non-Gaussian operations: Conditional photon subtraction versus cubic phase gate}",
    eprint = "2210.04531",
    archivePrefix = "arXiv",
    primaryClass = "quant-ph",
    doi = "10.1103/PhysRevA.107.022422",
    journal = "Phys. Rev. A",
    volume = "107",
    number = "2",
    pages = "022422",
    year = "2023"
}

@article{tele3,
    author = "Zinatullin, E. R. and Korolev, S. B. and Golubeva, T. Yu.",
    title = "{Teleportation with a cubic phase gate}",
    eprint = "2107.02511",
    archivePrefix = "arXiv",
    primaryClass = "quant-ph",
    doi = "10.1103/PhysRevA.104.032420",
    journal = "Phys. Rev. A",
    volume = "104",
    pages = "032420",
    year = "2021"
}

@article{kript1,
    author = "Guo, Ying and Ye, Wei and Zhong, Hai and Liao, Qin",
    title = "{Continuous-variable quantum key distribution with non-Gaussian quantum catalysis}",
    eprint = "1811.06698",
    archivePrefix = "arXiv",
    primaryClass = "quant-ph",
    doi = "10.1103/PhysRevA.99.032327",
    journal = "Phys. Rev. A",
    volume = "99",
    number = "3",
    pages = "032327",
    year = "2019"
}

@article{kript2,
    author = "Lee, Jaehak and Park, Jiyong and Nha, Hyunchul",
    title = "{Quantum non-Gaussianity and secure quantum communication}",
    eprint = "1907.05222",
    archivePrefix = "arXiv",
    primaryClass = "quant-ph",
    doi = "10.1038/s41534-019-0164-9",
    journal = "npj Quantum Inf.",
    volume = "5",
    pages = "49",
    year = "2019"
}

@article{kor1,
    author = "Hastrup, Jacob and Andersen, Ulrik Lund",
    title = "{Analysis of loss correction with the Gottesman-Kitaev-Preskill code}",
    eprint = "2112.01425",
    archivePrefix = "arXiv",
    primaryClass = "quant-ph",
    doi = "10.1103/PhysRevA.108.052413",
    journal = "Phys. Rev. A",
    volume = "108",
    number = "5",
    pages = "052413",
    year = "2023"
}

@article{kor2,
	author = {Korolev, S. B. and Bashmakova, E. N. and Golubeva, T. Yu.},
	date = {2024/10/18},
	doi = {10.1007/s11128-024-04549-w},
	id = {Korolev2024},
	isbn = {1573-1332},
	journal = {Quantum Information Processing},
	number = {10},
	pages = {354},
	title = {Error correction using squeezed Fock states},
	url = {https://doi.org/10.1007/s11128-024-04549-w},
	volume = {23},
	year = {2024}}

@article{Korolev_2024,
doi = {10.1088/1612-202X/ad6e6f},
url = {https://doi.org/10.1088/1612-202X/ad6e6f},
year = {2024},
month = {aug},
publisher = {IOP Publishing},
volume = {21},
number = {9},
pages = {095204},
author = {Korolev, S B and Bashmakova, E N and Yu Golubeva, T},
title = {Estimation of the set of states obtained in particle number measurement schemes},
journal = {Laser Physics Letters}
}

@article{kor3,
    author = "Bashmakova, E. N. and Korolev, S. B. and Golubeva, T. Yu.",
    title = "{Bosonic quantum error correction using squeezed Fock states}",
    eprint = "2506.00300",
    archivePrefix = "arXiv",
    primaryClass = "quant-ph",
    doi = "10.1103/97yt-nzg2",
    journal = "Phys. Rev. A",
    volume = "112",
    number = "3",
    pages = "032434",
    year = "2025"
}

@article{kor4,
    author = "Michael, Marios H. and Silveri, Matti and Brierley, R. {\,}T. and Albert, Victor V. and Salmilehto, Juha and Jiang, Liang and Girvin, S. {\,}M.",
    title = "{New Class of Quantum Error-Correcting Codes for a Bosonic Mode}",
    eprint = "1602.00008",
    archivePrefix = "arXiv",
    primaryClass = "quant-ph",
    doi = "10.1103/PhysRevX.6.031006",
    journal = "Phys. Rev. X",
    volume = "6",
    number = "3",
    pages = "031006",
    year = "2016"
}

@article{kor5,
    author = "Korolev, S. B. and Golubeva, T. Yu.",
    title = "{Bosonic error correction codes based on states generated via particle-number-resolving measurements}",
    eprint = "2509.16993",
    archivePrefix = "arXiv",
    primaryClass = "quant-ph",
    doi = "10.1103/hyk9-zxt2",
    journal = "Phys. Rev. A",
    volume = "113",
    number = "2",
    pages = "022413",
    year = "2026"
}

@article{otk1,
    author = "Mirrahimi, Mazyar and Leghtas, Zaki and Albert, Victor V. and Touzard, Steven and Schoelkopf, Robert J. and Jiang, Liang and Devoret, Michel H.",
    title = "{Dynamically protected cat-qubits: a new paradigm for universal quantum computation}",
    eprint = "1312.2017",
    archivePrefix = "arXiv",
    primaryClass = "quant-ph",
    doi = "10.1088/1367-2630/16/4/045014",
    journal = "New J. Phys.",
    volume = "16",
    number = "4",
    pages = "045014",
    year = "2014"
}

@article{otk2,
    author = "Schlegel, David S. and Minganti, Fabrizio and Savona, Vincenzo",
    title = {{Quantum error correction using squeezed Schr{\"o}dinger cat states}},
    eprint = "2201.02570",
    archivePrefix = "arXiv",
    primaryClass = "quant-ph",
    doi = "10.1103/PhysRevA.106.022431",
    journal = "Phys. Rev. A",
    volume = "106",
    number = "2",
    pages = "022431",
    year = "2022"
}

@article{los1,
    author = "Aspelmeyer, Markus and Kippenberg, Tobias J. and Marquardt, Florian",
    title = "{Cavity Optomechanics}",
    eprint = "1303.0733",
    archivePrefix = "arXiv",
    primaryClass = "cond-mat.mes-hall",
    doi = "10.1103/RevModPhys.86.1391",
    journal = "Rev. Mod. Phys.",
    volume = "86",
    pages = "1391",
    year = "2014"
}

@article{los2,
    author = "Gu, Xiu and Kockum, Anton Frisk and Miranowicz, Adam and Liu, Yu-xi and Nori, Franco",
    title = "{Microwave photonics with superconducting quantum circuits}",
    eprint = "1707.02046",
    archivePrefix = "arXiv",
    primaryClass = "quant-ph",
    doi = "10.1016/j.physrep.2017.10.002",
    journal = "Phys. Rept.",
    volume = "718-719",
    pages = "1--102",
    year = "2017"
}

@article{los3,
    author = "Schlegel, David S. and Minganti, Fabrizio and Savona, Vincenzo",
    title = {{Quantum error correction using squeezed Schr{\"o}dinger cat states}},
    eprint = "2201.02570",
    archivePrefix = "arXiv",
    primaryClass = "quant-ph",
    doi = "10.1103/PhysRevA.106.022431",
    journal = "Phys. Rev. A",
    volume = "106",
    number = "2",
    pages = "022431",
    year = "2022"
}

@article{Jeong_2015,
   title={Characterizations and quantifications of macroscopic quantumness and its implementations using optical fields},
   volume={337},
   ISSN={0030-4018},
   url={http://dx.doi.org/10.1016/j.optcom.2014.07.012},
   DOI={10.1016/j.optcom.2014.07.012},
   journal={Optics Communications},
   publisher={Elsevier BV},
   author={Jeong, Hyunseok and Kang, Minsu and Kwon, Hyukjoon},
   year={2015},
   month=Feb, pages={12–21} }

@article{Glauber1963,
  author = {Glauber, R. J.},
  title = {Coherent and Incoherent States of the Radiation Field},
  journal = {Physical Review},
  volume = {131},
  number = {6},
  pages = {2766--2788},
  year = {1963},
  publisher = {American Physical Society},
  doi = {10.1103/PhysRev.131.2766}
}

@article{DODONOV1974597,
title = {Even and odd coherent states and excitations of a singular oscillator},
journal = {Physica},
volume = {72},
number = {3},
pages = {597-615},
year = {1974},
issn = {0031-8914},
doi = {https://doi.org/10.1016/0031-8914(74)90215-8},
url = {https://www.sciencedirect.com/science/article/pii/0031891474902158},
author = {V.V. Dodonov and I.A. Malkin and V.I. Man'ko}
}

@article{PhysRevLett.57.13,
  title = {Generating quantum mechanical superpositions of macroscopically distinguishable states via amplitude dispersion},
  author = {Yurke, B. and Stoler, D.},
  journal = {Phys. Rev. Lett.},
  volume = {57},
  issue = {1},
  pages = {13--16},
  numpages = {0},
  year = {1986},
  month = {Jul},
  publisher = {American Physical Society},
  doi = {10.1103/PhysRevLett.57.13},
  url = {https://link.aps.org/doi/10.1103/PhysRevLett.57.13}
}

@article{PhysRevA.73.023803,
  title = {Sub-Planck phase-space structures and Heisenberg-limited measurements},
  author = {Toscano, F. and Dalvit, D. A. R. and Davidovich, L. and Zurek, W. H.},
  journal = {Phys. Rev. A},
  volume = {73},
  issue = {2},
  pages = {023803},
  numpages = {7},
  year = {2006},
  month = {Feb},
  publisher = {American Physical Society},
  doi = {10.1103/PhysRevA.73.023803},
  url = {https://link.aps.org/doi/10.1103/PhysRevA.73.023803}
}

@article{Dalvit_2006,
doi = {10.1088/1367-2630/8/11/276},
url = {https://doi.org/10.1088/1367-2630/8/11/276},
year = {2006},
month = {nov},
publisher = {},
volume = {8},
number = {11},
pages = {276},
author = {Dalvit, D A R and de Matos Filho, R L and Toscano, F},
title = {Quantum metrology at the Heisenberg limit with ion trap motional compass states},
journal = {New Journal of Physics}
}

@article{PhysRevA.106.043704,
  title = {Sub-Planck phase-space structure and sensitivity for SU(1,1) compass states},
  author = {Akhtar, Naeem and Sanders, Barry C. and Xianlong, Gao},
  journal = {Phys. Rev. A},
  volume = {106},
  issue = {4},
  pages = {043704},
  numpages = {14},
  year = {2022},
  month = {Oct},
  publisher = {American Physical Society},
  doi = {10.1103/PhysRevA.106.043704},
  url = {https://link.aps.org/doi/10.1103/PhysRevA.106.043704}
}

@article{PhysRevA.108.043719,
  title = {Superposing compass states for asymptotic isotropic sub-Planck phase-space sensitivity},
  author = {Shukla, Atharva and Sanders, Barry C.},
  journal = {Phys. Rev. A},
  volume = {108},
  issue = {4},
  pages = {043719},
  numpages = {14},
  year = {2023},
  month = {Oct},
  publisher = {American Physical Society},
  doi = {10.1103/PhysRevA.108.043719},
  url = {https://link.aps.org/doi/10.1103/PhysRevA.108.043719}
}

@article{Zurek2001,
  author = {Zurek, W. H.},
  title = {Sub-Planck structure in phase space and its relevance for quantum decoherence},
  journal = {Nature},
  volume = {412},
  number = {6848},
  pages = {712--717},
  year = {2001},
  publisher = {Nature Publishing Group},
  doi = {10.1038/35089017}
}

@article{PhysRevA.103.053711,
  title = {Sub-Planck structures: Analogies between the Heisenberg-Weyl and SU(2) groups},
  author = {Akhtar, Naeem and Sanders, Barry C. and Navarrete-Benlloch, Carlos},
  journal = {Phys. Rev. A},
  volume = {103},
  issue = {5},
  pages = {053711},
  numpages = {14},
  year = {2021},
  month = {May},
  publisher = {American Physical Society},
  doi = {10.1103/PhysRevA.103.053711},
  url = {https://link.aps.org/doi/10.1103/PhysRevA.103.053711}
}

@article{Lan_2022,
   title={Multi-Headed Symmetrical Superpositions of Coherent States},
   volume={61},
   ISSN={1572-9575},
   url={http://dx.doi.org/10.1007/s10773-022-05134-6},
   DOI={10.1007/s10773-022-05134-6},
   number={5},
   journal={International Journal of Theoretical Physics},
   publisher={Springer Science and Business Media LLC},
   author={Lan, Bo and Xu, Xue-xiang},
   year={2022},
   month=May }

@article{Korolev_fgen,
  title = {Generation of squeezed Fock states by measurement},
  author = {Korolev, S. B. and Bashmakova, E. N. and Tagantsev, A. K. and Golubeva, T. Yu.},
  journal = {Phys. Rev. A},
  volume = {109},
  issue = {5},
  pages = {052428},
  numpages = {8},
  year = {2024},
  month = {May},
  publisher = {American Physical Society},
  doi = {10.1103/PhysRevA.109.052428},
  url = {https://link.aps.org/doi/10.1103/PhysRevA.109.052428}
}

@Article{app142311249,
AUTHOR = {Adam, Peter and Mechler, Matyas},
TITLE = {Recent Progress in Multiplexed Single-Photon Sources},
JOURNAL = {Applied Sciences},
VOLUME = {14},
YEAR = {2024},
NUMBER = {23},
ARTICLE-NUMBER = {11249},
URL = {https://www.mdpi.com/2076-3417/14/23/11249},
ISSN = {2076-3417},
DOI = {10.3390/app142311249}
}

@article{DODONOV2016296,
title = {Decoherence of odd compass states in the phase-sensitive amplifying/dissipating environment},
journal = {Annals of Physics},
volume = {371},
pages = {296-312},
year = {2016},
issn = {0003-4916},
doi = {https://doi.org/10.1016/j.aop.2016.04.019},
url = {https://www.sciencedirect.com/science/article/pii/S0003491616300434},
author = {V.V. Dodonov and C. Valverde and L.S. Souza and B. Baseia}
}

@article{deFreitas_2021,
  title = {Non-Gaussianity of Four-Photon Superpositions of Fock States},
  volume = {3},
  ISSN = {2624-960X},
  url = {http://dx.doi.org/10.3390/quantum3030022},
  DOI = {10.3390/quantum3030022},
  number = {3},
  journal = {Quantum Reports},
  publisher = {MDPI AG},
  author = {de Freitas,  Miguel Citeli and Dodonov,  Viktor V.},
  year = {2021},
  month = July,
  pages = {350–365}
}

@article{Fiurasek_2026_1,
  title = {Maximum heralding probabilities of nonclassical-state generation from a two-mode Gaussian state via photon-counting measurements},
  author = {Fiur\'a\ifmmode \check{s}\else \v{s}\fi{}ek, Jarom\'{\i}r},
  journal = {Phys. Rev. A},
  volume = {113},
  issue = {6},
  pages = {063716},
  numpages = {11},
  year = {2026},
  month = {Jun},
  publisher = {American Physical Society},
  doi = {10.1103/nq5h-1rjf},
  url = {https://link.aps.org/doi/10.1103/nq5h-1rjf}
}

@article{Fiurasek_2026_2,
doi = {10.1088/2058-9565/ae9185},
url = {https://doi.org/10.1088/2058-9565/ae9185},
year = {2026},
month = {aug},
publisher = {IOP Publishing},
volume = {11},
number = {3},
pages = {035074},
author = {Fiurášek, Jaromír},
title = {Heralding probability optimization for nonclassical light generated by photon counting measurements on multimode Gaussian states},
journal = {Quantum Science and Technology}
}

\appendix
\section{Solution} \label{append_solut}
 Solution for the parameters of the four-mode Gaussian state used to generate 4-states

\begin{align} 
    &a_1 = \frac{(4 + c_1^2)(1 - c_5^2 c_3)(1 + c_3) + (c_5^2(1 + 2c_3) - 1)c_1 c_2 c_4 + (1 + c_5^2)c_4^2 - c_2^2(c_5^2(c_4^2 + 1) + 1)}
{(4 - c_1^2)(1 - c_5^2 c_3)(1 + c_3) - c_1 c_2(c_5^2(1 + 2c_3) - 1)c_4 - (1 + c_5^2)c_4^2 + c_2^2(c_5^2(c_4^2 - 1) - 1)}, \label{a1}\\
& a_2 = \frac{(4 + c_1^2)(1 - c_5^2 c_3)(1 + c_3) + (c_5^2(1 + 2c_3) - 1)c_1 c_2 c_4 - (1 + c_5^2)c_4^2 + c_2^2(1 - c_5^2(c_4^2 - 1))}
{(4 - c_1^2)(1 - c_5^2 c_3)(1 + c_3) - (c_5^2(1 + 2c_3) - 1)c_1 c_2 c_4 - (1 + c_5^2)c_4^2 + c_2^2(c_5^2(c_4^2 - 1) - 1)},\\
& a_3 = -\frac{(4 - c_1^2)(1 - c_5^2 c_3)(c_3 - 1) + (c_5^2(1 - 2c_3) + 1)c_1 c_2 c_4 + (c_5^2 - 1)c_4^2 + c_2^2(c_5^2(1 + c_4^2) - 1)}
{(4 - c_1^2)(1 - c_5^2 c_3)(1 + c_3) - (c_5^2(1 + 2c_3) - 1)c_1 c_2 c_4 - (1 + c_5^2)c_4^2 + c_2^2(c_5^2(c_4^2 - 1) - 1)},\\
& a_4 = \frac{(4 - c_1^2)(1 + c_5^2 c_3)(1 + c_3) + (c_5^2(1 + 2c_3) + 1)c_1 c_2 c_4 + (c_5^2 - 1)c_4^2 - c_2^2(c_5^2(c_4^2 - 1) + 1)}
{(4 - c_1^2)(1 - c_5^2 c_3)(1 + c_3) - c_1 c_2(c_5^2(1 + 2c_3) - 1)c_4 - (c_5^2 + 1)c_4^2 + c_2^2(c_5^2(c_4^2 - 1) - 1)},\\
& b_{12} = \frac{2\bigl(2c_1(1 + c_3)(c_5^2 c_3 - 1) - c_2(c_5^2(1 + 2c_3) - 1)c_4\bigr)}
{(4 - c_1^2)(1 - c_5^2 c_3)(1 + c_3) - c_1 c_2(c_5^2(1 + 2c_3) - 1)c_4 - (1 + c_5^2)c_4^2 + c_2^2(c_5^2(c_4^2 - 1) - 1)},\\
& b_{13} = \frac{2\bigl(c_1 c_2(1 - c_5^2 c_3) + (c_5^2(c_2^2 + 2c_3) - 2)c_4\bigr)}
{(4 - c_1^2)(1 - c_5^2 c_3)(1 + c_3) - c_1 c_2(c_5^2(1 + 2c_3) - 1)c_4 - (1 + c_5^2)c_4^2 + c_2^2(c_5^2(c_4^2 - 1) - 1)},\\
& b_{14} = \frac{2\bigl(c_1 c_2 c_5(1 + c_3) - c_5(c_2^2 - 2(1 + c_3))c_4\bigr)}
{(4 - c_1^2)(1 - c_5^2 c_3)(1 + c_3) - c_1 c_2(c_5^2(1 + 2c_3) - 1)c_4 - (1 + c_5^2)c_4^2 + c_2^2(c_5^2(c_4^2 - 1) - 1)},\\
& b_{23} = \frac{2\bigl(c_1(1 - c_5^2 c_3)c_4 + c_2(c_5^2(2c_3 + c_4^2) - 2)\bigr)}
{(4 - c_1^2)(1 - c_5^2 c_3)(1 + c_3) - c_1 c_2(c_5^2(1 + 2c_3) - 1)c_4 - (1 + c_5^2)c_4^2 + c_2^2(c_5^2(c_4^2 - 1) - 1)},\\
&b_{24} = \frac{2c_6\bigl(-c_1(1 + c_3)c_4 + c_2(c_4^2 - 2(1 + c_3))\bigr)}
{(4 - c_1^2)(1 - c_5^2 c_3)(1 + c_3) - c_1 c_2(c_5^2(1 + 2c_3) - 1)c_4 - (1 + c_5^2)c_4^2 + c_2^2(c_5^2(c_4^2 - 1) - 1)},\\
& b_{34} = \frac{2c_6(c_2 - c_4)(c_2 + c_4)}
{(4 - c_1^2)(1 - c_5^2 c_3)(1 + c_3) - c_1 c_2(c_5^2(1 + 2c_3) - 1)c_4 - (1 + c_5^2)c_4^2 + c_2^2(c_5^2(c_4^2 - 1) - 1)} \label{b34}.
\end{align}

\section{Distribution of the amplitudes of the compass state} \label{append_distrib_coef}
\begin{figure}[H]
    \centering
    \includegraphics[width=0.5\linewidth]{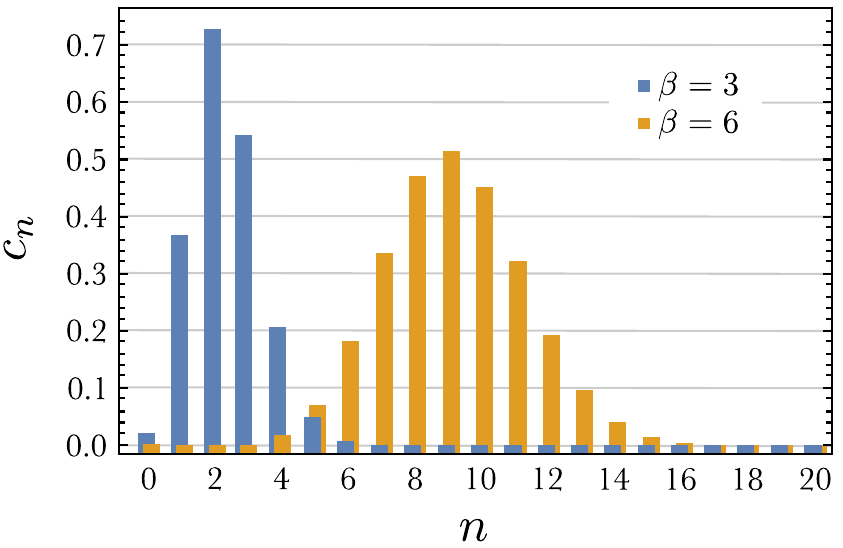}
    \caption{Distribution of the expansion coefficients of the compass state over Fock states.}
    \label{fig:distribution}
\end{figure}

Fig. \ref{fig:distribution} shows the distribution of the expansion coefficients of the compass state (\ref{eq:Phi_k}) over Fock states for two amplitudes: $\beta = 3$ and $\beta = 6$.

\section{Expansion coefficients of the 8-states} \label{append_decompos_coef_8}
Expansion coefficients of the states generated in the scheme shown in Fig.~\ref{fig:scheme_3} over Fock states,

\begin{multline}
  \gamma_j (A,B)=  \left(\frac{A^2}{4}\right)^{2 n-j} \sum _{j_1=0}^{j_{max}} \sum _{j_2=0}^{j_{max}} (-1)^{j_2}\sqrt{\frac{\left(m-2 n+4 j_2\right)!}{\left(m-2 n+4 j_1\right)! (2 m+4n-k-4 j)!}} \times\\
  \left(2 j_1\right)! \left(2 j_2\right)! \left(2 n-2 j_1\right)! \left(2 n-2 j_2\right)! \binom{n}{n-j_1} \binom{n}{n-j_2}\times\\
  \, _2\tilde{F}_1\left(-m,2 n-m-4 j_1;2 n-m-2 j_1+1;B\right) \, _2\tilde{F}_1\left(-m,2 n-m-4 j_2;2 n-m-2
   j_2+1;B\right)\times\\
   \delta _{2 n-j,j_1+j_2} \, _2\tilde{F}_1\left(k-2 m-4 n+4 j,2 n-m-4 j_1;k-m-6 n+4 j+4 j_2+1;-1\right),
\end{multline}
where $\delta_{2n-j,\, j_1+j_2}$ is the Kronecker delta, and the upper summation limit is given by $j_{\max} = \max\left(0, \left\lceil \frac{1}{4}(2n-m) \right\rceil \right)$.

\end{document}